\documentclass[reprint,
 aps,
]{revtex4-2}

\usepackage{hyperref}
\usepackage{graphicx}
\usepackage{dcolumn}
\usepackage{bm}

\usepackage{amsmath,amssymb,amsthm,easybmat,verbatim}
\usepackage{enumerate}
\usepackage{color}
\usepackage{bm}
\usepackage{graphicx}
\usepackage{longtable}
\usepackage{array}
\usepackage{tikz}
\usepackage{float}

\newtheorem{corollary}{Corollary}
\newtheorem{definition}{Definition}

\newtheorem{lemma}{Lemma}

\newtheorem{theorem}{Theorem}

\newtheorem*{remark}{Remark}

\usepackage{physics}

\newcommand{\beq}{\begin{equation}}
\newcommand{\eeq}{\end{equation}}
\newcommand{\bea}{\begin{eqnarray}}
\newcommand{\eea}{\end{eqnarray}}

\newcommand{\kett}[1]{|#1\rangle}
\newcommand{\brat}[1]{\langle#1|}

\begin{document}


\title{Sufficient Conditions and Constraints for Reversing General Quantum Errors}

\author{Alvin Gonzales$^{2}$}
\email{agonza@siu.edu}

\author{Daniel Dilley$^1$}

\author{Mark Byrd$^{1,2}$}

\affiliation{$^1$Department of Physics, Southern Illinois University Carbondale, Carbondale, Illinois 62901, USA}
\affiliation{$^2$School of Computing, Southern Illinois University Carbondale, Carbondale, Illinois 62901, USA}

\date{\today}

\begin{abstract}
Reversing the effects of a quantum evolution, for example as is done in error correction, is an important task for controlling quantum systems in order to produce reliable quantum devices.  When the evolution is governed by a completely positive map, there exist reversibility conditions, known as the quantum error correcting code conditions, which are necessary and sufficient conditions for the reversibility of a quantum operation on a subspace, the code space.  However, if we suppose that the evolution is not described by a completely positive map, necessary and sufficient conditions are not known.  Here we consider evolutions that 
do not necessarily correspond to a completely positive map.  We prove that the completely positive map error correcting code conditions can lead to a code space that is not in the domain of the map, meaning that the output of the map is not positive.  A corollary to our theorem provides a class of relevant examples.  Finally, we provide a set of sufficient conditions that will enable the use of quantum error correcting code conditions while ensuring positivity.   
\end{abstract}

\maketitle


\section{Introduction}

Reversing quantum operations is an important form of quantum control which will help to enable many quantum technologies.  For example, error correction, which is the reversal of an unwanted quantum operation, will be necessary to ensure that errors do not ruin a quantum computer's algorithm execution.  Error correction is also important in long-distance communication to ensure data integrity.  Quantum error correction was shown to be possible, and potentially practical, with the invention of the Shor \cite{shor_1995} and Steane \cite{Steane:96a} quantum error correcting (QEC) codes.  Subsequently, with a set of reasonable assumptions, necessary and sufficient conditions for the existence of an error correcting code were provided by Bennett \textit{et al}. \cite{bennett_1996}, Knill and Laflamme \cite{knill_1997}, and Nielsen \textit{et al}. \cite{Nielsen_Caves_Schumacher_Barnum_1998}.  While these conditions are for exact recovery, it is possible to approximately reverse or recover a state \cite{leung1997ApproxQEC, beny2010ApproxQEC}.

Such conditions are often described in terms of a completely positive (CP) map ${\mathcal A}$, that is, a mapping that takes all positive operators to positive operators and does so even when extended by an identity operator to $I_n\otimes {\mathcal A}$, where $I_n$ is the $n\times n$ identity operator.  This is sometimes also called a dynamical map, although not all maps are completely positive (e.g., the transpose) and the terminology is not consistent in literature with respect to dynamical maps. It should also be noted that there is an ongoing discussion in the physics community about the physicality of noncompletely positive (NCP) maps.  (See, for example, \cite{Schmid_Ried_Spekkens_2018} and references therein.)  However, most researchers consider a map physical if the domain of the map is restricted to positive output density operators  \cite{Jordan_Shaji_Sudarshan_2004, Shaji_Sudarshan_2005}. 

Without directly addressing the physicality of the map here, we present conditions which restrict the ability to reverse an evolution that does not correspond to a CP map.  Normally, one needs to carefully consider if the map is ``physical" (i.e., it gives an accurate description of the physical process). A single system evolving in time from one state to another is generally not enough to define a physical map \cite{Schmid_Ried_Spekkens_2018}. However, in this paper we specifically focus on reversing maps $\mathcal{E}$ as defined in our Eq.~\eqref{eq:ncpevol} below and we do \textit{not} restrict $\mathcal{E}$ to correspond to a physical map. Instead, we assume that $\mathcal{E}$ gives the observed/measured final evolution for states in its domain. The objective is to reverse the effect of the evolution of the state and  whether the map describing that evolution is physical or not does not change our results. The results apply whether or not the map describing the system evolution is ``physical."  Note that if only a single input and output of an evolution are given, one can always find a CP map that corresponds to this evolution \cite{Nielsen_Chuang_Textbook_2011}.  However, if other information is given, this may not be the case \cite{Chitambaretal:2015}.

When we study the reversibility of a system, particularly for error correction, we are often looking at a subspace $\mathcal{H}_{S^\prime}$ of the system-environment Hilbert space $\mathcal{H}_{SE}$, where $E$ is the environment, $S$ is the system, and $\mathcal{H}_{S^\prime} \subset \mathcal{H}_{S}$. The initial state of the system is $\rho_S=\tr_E(\rho_{SE})$, where $\rho_{SE}$ is the initial combined system and environment state.  One can in principle experimentally determine a dynamical map $\mathcal{A}:\rho_S\rightarrow \rho_S^\prime$ that describes the open-system evolution of the system under consideration.  This will determine the set of errors that occurs on the system, and an appropriate error-correcting code can be determined from the set of errors that is targeted for correction.  It is well known that when the initial state of the system and environment together is a product state, that is, when they are uncorrelated, $\rho_{SE}=\rho_S\otimes \rho_E$ and the evolution can be described by a CP map.  

\begin{figure}[H]
    \centering
    \includegraphics[scale=1]{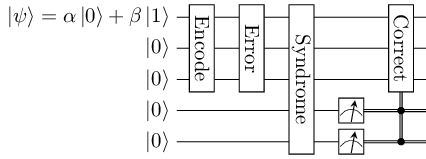}
    \caption{Single bit flip-error correction process.\\
    The measurement is part of the syndrome extraction process, but is drawn separately here for emphasis.}
    \label{fig:singleBitFlip}
\end{figure}

The model of error correction that we consider is where the recovery operation is implemented after the error.  This is the model usually considered and is, for example, discussed in some detail in the book by Nielsen and Chuang \cite{Nielsen_Chuang_Textbook_2011}. 
To be more specific, the process for quantum error correction occurs in four main steps.  In the first step, the system is encoded. Next, the system evolves, possibly incurring an error. Then, a measurement is made to extract the error syndrome to identify a possible correctable error. Finally, the error, if present, is corrected using a unitary transformation.  The way to express this, arising from Eq.~(\ref{eq:cpecc}) (below), is $U_kP_k$, where $P_k$ is the measurement to detect an error and $U_k$ is the corresponding unitary which is implemented conditioned on the outcome of the measurement $P_k$.  A diagram of an example of this process for the single bit-flip repetition code is shown in Fig. \ref{fig:singleBitFlip}. The details of the gates in Fig. \ref{fig:singleBitFlip} are not important for our situation.  However, it should be emphasized that the correction process depends on the syndrome measurement outcome.

When the evolution does not correspond to a CP map, we would like to find a way to generalize or extend the reversibility conditions [CP error correcting conditions \eqref{eq:cpecc}].  Motivated by a desire to describe very general error models such as those considered by Aharonov and Ben-Or \cite{aharonov2008FaultTolQuantComp},  Shabani and Lidar \cite{shabani_2009} studied this problem and showed that the same code space for the corresponding CP map works for a corresponding NCP map (specified below), but they focused on the Hermiticity of the evolution and not the positivity.  

In this paper, we show that if an evolution is not described by a CP map, satisfying the CP quantum error correcting conditions can produce a code space that is not in the domain of the NCP error map in the sense that it does not produce a positive output.  In contrast to \cite{shabani_2009}, we seek an output that is not only Hermitian but also positive.  We provide conditions on the code, via Theorem 1, such that the quantum error correcting conditions for a NCP map will produce a nonpositive, Hermitian output.  This leads to a set of sufficient conditions for the reversibility of a NCP map when we demand that the output be both Hermitian and positive.  The conditions in Ref.~\cite{shabani_2009} are sufficient only if the positivity is not in question.  This is stated in Theorem 2 and followed by instructive examples.   


\section{Background}

A superoperator $\mathcal{A}$ can be represented by a matrix acting on $\rho_S$ \cite{sudarshan_1961}:
\begin{align}
    \rho'_{r',s'}=\mathcal{A}_{r's',rs}\rho_{rs}.
\end{align}
(The sum over repeated indices is implied.)  The evolutions we consider will be those that preserve the Hermiticity and trace.  In this case, the matrix $\mathcal{A}$ must satisfy the conditions, respectively, 
\begin{align}
    &\mathcal{A}_{s'r',sr}=(\mathcal{A}_{r's',rs})^* \label{eq:Amatconds1}
\end{align}
and
\begin{align}\label{eq:Amatconds3}
    &\mathcal{A}_{r'r',rs}=\delta_{rs},
\end{align}
where $^*$ is the complex conjugate.

For an alternative description, we often use the matrix $\mathcal{B}$, which is related to the $\mathcal{A}$ matrix by
\begin{align}
    \mathcal{B}_{r'r,s's}=\mathcal{A}_{r's',rs}.
\end{align}
The Hermiticity condition \eqref{eq:Amatconds1} translates to
\begin{align}
    &\mathcal{B}_{r'r,s's}=(\mathcal{B}_{s's,r'r})^*.\label{eq:Bmatconds1}
\end{align}
Then a general Hermitian preserving linear map can be written in an operator-sum decomposition $\mathcal{E}$ of the form
\begin{align}\label{eq:ncpevol}
    \mathcal{E}(\rho)=\sum_i\eta_iE_i\rho E^\dagger_i,
\end{align}
where the $\{\eta_i\}$'s are the signs of the eigenvalues, and $\{E_i\}$'s are the eigenvectors of the matrix $\mathcal{B}$ after absorbing the magnitudes of the eigenvalues \cite{sudarshan_1961,Choi_1975, Jordan_Shaji_Sudarshan_2004}.  The eigenvectors are written in matrix form.  

If the system is not correlated with the environment, i.e., the combined system and environment is a product state $\rho_S\otimes \rho_E$, then the evolution of the system is given by a completely positive map, 
and all the $\eta_i=1$. In the case that the system and environment are not initially in a product state, general conditions for complete positivity are not known, but in some special cases the map is still CP \cite{Rodriguez-Rosario_2008, Carteret_Terno_2008, shabani2009VanishQuantumDiscord, shabani2016erratum, modi2012PositivityInThePresOfInitSECorr, brodutch2013VanishingQuantumDiscord, buscemi2014CPMarkovianityAndTheQuanDatProc, liu2014CPMaps,Vacchini_Amato_2016}.  
However, whenever the map $\mathcal{E}(\rho)$ is CP, we can write it as \cite{Choi_1975, Kraus_1983} 
\begin{align}
    \mathcal{E}(\rho)=\sum_i E_i\rho E_i^\dagger.
\end{align}
Furthermore, when the evolution corresponds to a CP map, there is a set of quantum-error correcting code conditions, which ensures the reversibility of the evolution.  These are necessary and sufficient for the construction of a quantum error correcting code, which can be used to detect and correct the errors, thus reversing the effects of the map.  One way of expressing these conditions is \cite{knill_1997}
\beq
\bra{\alpha_L}E_i^\dagger E_j\ket{\beta_L} = m_{ij}\delta_{\alpha\beta},
\eeq
where $\ket{\alpha_L}$, $\ket{\beta_L}$ are logical (encoded states) and $m_{ij}$ is a constant.  

This equation is easy to interpret.  If $\ket{\alpha_L}$ is acted on by an error $E_i$ and $\ket{\beta_L}$ is another state acted on by an error $E_j$, then the overlap between these must be zero if the states are different.  This ensures that a measurement performed to identify the error will not result in an ambiguous correction procedure to recover the original state.  This, and other manifestations in classical error correction, are sometimes called the ``disjointness condition,'' since it shows that the subspace of a logical state acted upon by any correctable error must be disjoint, as a set, from any other logical state with a correctable error acting on it.  It is easy to show that these conditions are satisfied if and only if the equivalent necessary and sufficient conditions for error correction for CP maps are satisfied \cite{Nielsen_Caves_Schumacher_Barnum_1998},
\begin{align}\label{eq:cpecc}
    PE_i^\dagger E_jP=c_{ij}P,
\end{align}
where $P$ is the projector onto the code space and $c_{ij}$ are elements of a Hermitian matrix.

A system can often develop correlations with its environment so that the combined system-environment state is no longer a product state, i.e., $\rho_{SE}\neq\rho_S\otimes \rho_E$. Correlations between the system and environment can be prevented with dynamical decoupling, but dynamical decoupling does not remove correlations that are present prior to the decoupling operations  \cite{viola_lloyd_1998,viola_knill_lloyd1999,Lidar_Brun:ECbook}. Given a correlated system and environment, the evolution of the system is not necessarily given by a CP map \cite{Pechukas_1994, Jordan_Shaji_Sudarshan_2004, Shaji_Sudarshan_2005, Rodriguez-Rosario_2008, Alicki_1995, Pechukas_1995}. A not completely positive evolution can be described by a $\mathcal{B}$ matrix that has at least one negative eigenvalue and has the operator-sum decomposition form
\begin{align}
    \mathcal{E}(\rho)=\sum_i\eta_iE_i\rho E^\dagger_i,
\end{align}
where the $\eta_i$'s are not all positive \cite{Jordan_Shaji_Sudarshan_2004}.  Such an evolution may not correspond to a physical map, but only a specific input and output state.


\section{Reversibility Conditions}

Our first main theorem shows that we need to be careful when extending results from CP maps to NCP maps if we want to ensure positivity.

First, let us define a pseudounitary (PU) transformation with signature $p,q$ to be a matrix $U$ such that $U\eta U^\dagger =\eta,$ and a pseudo-Hermitian (PH) matrix to be a matrix $H$ such that $H^\dagger = \eta H \eta^{-1}$, where, in our case, $\eta = \mbox{diag}(1,1,1...,1,-1,-1,...,-1)$ ($p$ ones and $q$ negative ones). (For a more general and thorough discussion, see \cite{Ali_Mostafazadeh_04} and Appendix \ref{sec:appendix1}.)  There exists a pseudounitary degree of freedom in the operator-sum decomposition that can be used to express a NCP map in terms of a different set of operators as shown in \cite{Ou_Byrd_2010}. (For completeness, we provide a slightly different proof in Appendix \ref{sec:appendix2} that we believe is clearer.)  The first task is to show that the matrix $c_{ij}$ in Lemma \ref{lemma:diag} can be diagonalized by choosing a pseudounitary transformation which will transform $E_i$ to a new set that produces the same map but has $c_{ij}$ diagonal. We call Eq.~\eqref{eq:ncpecc0} (below) the pseudo-Hermitian form of the CP error correcting conditions because when you diagonalize it, you get the diagonalized CP error correcting conditions. 
\begin{lemma}\label{lemma:diag}
    Given a NCP map $\mathcal{E}(\rho)=\sum_i\eta_iE_i\rho E_i^\dagger,$ the PH form of the CP error correcting conditions
    \begin{align}\label{eq:ncpecc0}
        \eta_i P E_i^\dagger E_jP = c_{ij}P,
    \end{align}
    where $c_{ij}$ are elements of a pseudo-Hermitian matrix $C$, can be diagonalized using the pseudounitary degree of freedom, and it leads to the diagonalized CP error correcting conditions.
\end{lemma}
\begin{proof}
We can choose a PU transformation $U$ with elements $u_{kj}$ such that $F_j=E_ku_{kj}$.  In other words, the $F_j$ are linear combinations of the $E_k$ with the set of coefficients $u_{kj}$ forming a PU matrix.  

We can diagonalize $c_{ij}$ in \eqref{eq:ncpecc0} by using the pseudounitary degree of freedom of the operators. We switch to block matrix notation by letting
\begin{align}\label{eq:blockmat0}
    &F=\begin{bmatrix}
        F_1 & F_2 & \dots & F_n
    \end{bmatrix},\\
    \label{eq:blockmat1}&E=\begin{bmatrix}
        E_1 & E_2 & \dots & E_n
    \end{bmatrix},\\
    \label{eq:blockmat2}&P=\mathbb{I}_n\otimes P,
\end{align}
where in \eqref{eq:blockmat2}, $\mathbb{I}_n$ is the $n\times n$ identity matrix. We can make the number of elements equal in $F$ and $E$ by inserting zero matrices. We treat the block components in \eqref{eq:blockmat0}, \eqref{eq:blockmat1}, and \eqref{eq:blockmat2} as elements so $F=EU$, $UP=PU$, and $\eta P=P\eta$. Letting  $M=E^\dagger E$, we have
\begin{align}
\label{eq:diagph1}\eta PF^\dagger FP&=\eta PU^\dagger MUP\\
\label{eq:diagph2}&=P\eta U^\dagger\eta\eta MUP\\
\label{eq:diagph3}&=PU^{-1}\eta MUP\\
\label{eq:diagph4}&=U^{-1}\eta PMPU\\
\label{eq:diagph5}&=U^{-1}CUP\\
\label{eq:diagph6}&=DP,
\end{align}
where $D$ is diagonal pseudo-Hermitian and we used the property that a pseudounitary matrix can diagonalize a pseudo-Hermitian in \eqref{eq:diagph5} to get \eqref{eq:diagph6} (see the Appendix \ref{sec:appendix1} and \ref{sec:appendix2} for details). Since $D$ is diagonal, we have $D=\eta D^\dagger\eta=D^\dagger\eta\eta=D^\dagger$. Thus, $D$ is also Hermitian. Then we can bring $\eta$ to the right-hand side of \eqref{eq:diagph6} and absorb it into $D$ because $\eta D$ is also a diagonal Hermitian. We can simply write \eqref{eq:diagph6} as
\begin{align}\label{eq:cpec0}
PF^\dagger FP=DP,
\end{align}
where $D$ is a diagonal Hermitian, or in index notation
\begin{align}\label{eq:cpec1}
&PF^\dagger_iF_jP=d_{ij}\delta_{ij}P.
\end{align}
This is the same as the diagonalized quantum error correction conditions for CP maps.
\end{proof}
Note that the ability to diagonalize this matrix is tantamount to finding a set of orthogonal projectors that can be used to define a syndrome measurement.

We can now prove the main theorem.
\begin{theorem}\label{thm:nogo0}
Let us consider a $\mathcal{B}$ matrix with at least one negative eigenvalue.  Let its action correspond to a NCP map
\begin{align}
    \mathcal{E}(\rho)=\sum_i{\eta_i E_i\rho E_i^\dagger},
\end{align}
where not all of the $\eta_i$'s are positive. Now suppose $\exists U\in$PU relating two sets of operators $\{E_i\}$ and $\{F_j\}$ such that $\mathcal{E}$ is equivalent to
\begin{align}\label{eq:nogo1}
    \mathcal{E}(\rho)=\sum_j{\eta_j F_j\rho F_j^\dagger}=\mathcal{E}_1(\rho)-\mathcal{E}_2(\rho),
\end{align}
where $\mathcal{E}_1$ and $\mathcal{E}_2$ are both CP maps.  
In addition, assume that $PF_i^\dagger F_jP=Pd_{ij}\delta_{ij}$, i.e., we satisfy the diagonalized error correcting conditions. If $\mathcal{E}_2(P\rho P)\neq 0$, then the code space is not in the domain of the error map.  
\end{theorem}
\begin{proof}
Following Nielsen \textit{et al}. \cite{Nielsen_Caves_Schumacher_Barnum_1998}, we can show that enforcing the diagonal error correcting conditions leads to a code space that is not in the domain of the error map. Let our input density matrix be in the code space, which can be written as $P\rho P$.   

Starting from \eqref{eq:nogo1}, we can use the polar decomposition to get
$F_kP=U_k\sqrt{PF^\dagger_kF_kP}=\sqrt{d_{kk}}U_kP$. Therefore, $F_k$ rotates the code subspace into the subspace given by the projector 
\begin{align}
    P_k\equiv U_kPU^\dagger_k=F_kPU^\dagger_k/\sqrt{d_{kk}}.
\end{align}
Then
\begin{align}\label{eq:nogo2}
    F_kP=\sqrt{d_{kk}} P_kU_k.
\end{align}
The diagonal error correcting conditions ensure that these rotated subspaces are orthogonal, since when $k\neq l$,
\begin{align}\label{eq:nogo3}
    P_lP_k=P^\dagger_lP_k=\dfrac{U_lPF^\dagger_lF_kPU^\dagger_k}{\sqrt{d_{ll}d_{kk}}}=0.
\end{align}
Using \eqref{eq:nogo1}, \eqref{eq:nogo2}, and \eqref{eq:nogo3}, we can measure the output state and we have outcome $j$ (un-normalized):
\begin{align} \label{eq:nogo4}
    \notag \mathcal{M}_j(\rho')=&\mathcal{M}_j(\mathcal{E}(P\rho P))\\
    =&P_j\mathcal{E}_1(P\rho P)P_j 
    -P_j\mathcal{E}_2(P\rho P)P_j\\
    \notag=&\sum_{k\neq l}{\abs{d_{kk}}P_jP_kU_k\rho U_k^\dagger P_k P_j}\\
    &-\sum_{l\neq k}{\abs{d_{ll}}P_jP_lU_l\rho U_l^\dagger P_l P_j}.
\end{align}
Since $\mathcal{E}_2(P\rho P)\neq 0$, using the orthogonality of $P_i$ we can choose $j$ to be one of the $l$; thus, $P_jP_l=P_l$ for one $l$, and the other $P_i$ terms vanish.  Thus, the probability of the outcome is a negative value.  Since for any valid positive semidefinite density operator this is not possible, the code space cannot be in the domain of the error map.
\end{proof}

Theorem \ref{thm:nogo0} leads to a useful corollary. In Eq.~\eqref{eq:nogo4} we relied on the fact that $\mathcal{E}_2(P\rho P)\neq 0$. Thus, the negative terms are nonzero. It follows that if the $F$ operators are unitary, as is the case with the Pauli matrices, we also arrive at the restriction as shown in Corollary \ref{coll:ncpec0}.

\begin{corollary}\label{coll:ncpec0}
For a $\mathcal{B}$ matrix with at least one negative eigenvalue, if the pseudounitary degree of freedom leads to
\begin{align}\label{eq:ncpec1}
PF^\dagger_jF_iP=P\delta_{ji},
\end{align}
where the map operators $F_i$ are unitary and $P$ is the projector onto the code space, then the code space is not in the domain of the error map. 
\end{corollary}
\begin{proof}
The proof is similar to the proof for Theorem \ref{thm:nogo0}. Let
\begin{align}
    P_k\equiv F_kPF^\dagger_k.
\end{align}
Then
\begin{align}\label{eq:ncpec2.5}
    F_kP=P_kF_k.
\end{align}
Equation \eqref{eq:ncpec1} ensures that these rotated subspaces are orthogonal, since when $k\neq l$,
\begin{align}\label{eq:ncpec3}
    P_lP_k=F_lPF^\dagger_lF_kPF^\dagger_k=0.
\end{align}
 For states $P\rho P$ in the code space, we have $\mathcal{E}(P\rho P)=\sum_i\eta_iF_i P\rho PF_i^\dagger=\mathcal{E}_1(P\rho P)-\mathcal{E}_2(P\rho P)$. Using \eqref{eq:ncpec2.5} and \eqref{eq:ncpec3}, we measure the state and we have outcome $j$ (un-normalized):
\begin{align} \label{eq:ncpec4}
    \notag \mathcal{M}_j(\rho')=&\mathcal{M}_j(\mathcal{E}(P\rho P)) \\
    \notag =&P_j\mathcal{E}_1(P\rho P)P_j - P_j\mathcal{E}_2(P\rho P)P_j\\
    \notag =&\sum_{k\neq l} P_j F_kP\rho PF^\dagger_kP_j - \sum_{l\neq k} P_j F_lP\rho PF^\dagger_lP_j\\
    =&\sum_{k\neq l}{P_jP_kF_k\rho F_k^\dagger P_k P_j} - \sum_{l\neq k}{P_jP_lF_l\rho F_l^\dagger P_l P_j}.
\end{align}

Note that $\mathcal{E}_2(P\rho P)\neq 0$ because $F_lP\rho PF^\dagger_l\neq 0$, since $F_l$ is a unitary matrix and thus preserves rank. Using the orthogonality of the $P_i$ projectors, we can choose $j$ to be one of the $l$ in the map so that $P_jP_l=P_l$ for one $l$ and the other $P_i$ terms vanish. Thus, the probability of the outcome is a negative value. Therefore, enforcing \eqref{eq:ncpec1} results in a density matrix which has a negative eigenvalue, and the code space is not in the domain of $\mathcal{E}(\rho)$.
\end{proof}

Note, however, that if $\mathcal{E}_2(P\rho P)=0$, we can still satisfy the NCP error correcting conditions and our code space is in the domain of $\mathcal{E}$.  This is stated more formally in the following theorem.

\begin{theorem}\label{thm:suff}
Consider an evolution $\mathcal{E}(\rho)=\mathcal{E}_1(\rho)-\mathcal{E}_2(\rho)$ of the form \eqref{eq:nogo1}.  If the quantum error correcting code conditions \eqref{eq:cpec1} are satisfied, and ${\mathcal{E}}_2(P\rho P)=0$, i.e., the negative part of the map $\mathcal{E}(\rho)$ is zero on the code space, then the evolution can be reversed and the resulting density operator $\mathcal{E}(P\rho P)$ will be positive.  
\end{theorem}
\begin{proof}
The proof follows from Theorem 1 and the QEC code conditions for a CP map. Starting from \eqref{eq:nogo4}, we have (un-normalized)
\begin{align}
    \notag\mathcal{M}_j(\rho')&=P_j\mathcal{E}_1(P\rho P)P_j - P_j\mathcal{E}_2(P\rho P)P_j\\
    \notag &=P_j\mathcal{E}_1(P\rho P)P_j\\
    \notag &=\sum_{k}{\abs{d_{kk}}P_jP_kU_k\rho U_k^\dagger P_k P_j}.
\end{align}
From the orthogonality of $P_i$, we have $P_jP_k=P_k$ for one $k$ value, and the other $P_i$ terms vanish. The correction is finished by conjugating with $U_k^\dagger$ because $U_k^\dagger P_kU_k=P.$ This recovery process is given by the recovery map
\begin{align}
    \mathcal{R}(\rho)=\sum_j U_j^\dagger P_j\rho P_j U_j.
\end{align}
Since $\rho'=\mathcal{E}(P\rho P)=\mathcal{E}_1(P\rho P)$, $\rho'$ is clearly positive.
\end{proof}

\begin{remark}
We consider trace increasing maps to be nonphysical. Thus, we require that $\mathcal{E}$ is not a trace increasing map. It is important to note that Theorem \ref{thm:suff} does not violate this condition. This can be seen from the fact that if $\mathcal{E}(\rho)$ is not a trace increasing map, this condition holds for all $\rho$, including the subspace $P\rho P$. 
\end{remark}



\section{Examples}\label{sec:examples}

It is argued in Ref.~\cite{shabani_2009} that, given a NCP map $\Phi$, a corresponding CP map $\tilde{\Phi}$ can be defined by taking the absolute value of the coefficients in the operator-sum decomposition. Then, using this CP map, a code space and recovery map is determined, which also works for the original NCP map. According to our Theorem \ref{thm:nogo0}, this can lead to a non-positive outcome, which we show with an example. In Ref.~\cite{shabani_2009} it states
\begin{quote}
\textit{Corollary 1.}
Consider a Hermitian noise map $\Phi_{H}(\rho)=\sum_{i=1}^N c_iK_i\rho K_i^\dagger$ and associate to it a CP map $\tilde{\Phi}_{CP}(\rho)=\sum_{i=1}^N \abs{c_i}K_i\rho K_i^\dagger$. Then any QEC code $\mathcal{C}$ and corresponding CP recovery map $\mathcal{R}_{CP}$ for $\tilde{\Phi}_{CP}$ are also a QEC code and CP recovery map for $\Phi_H$.
\end{quote}
The following gives an example of when their \textit{Corollary 1} produces a nonpositive outcome which is covered by our Corollary \ref{coll:ncpec0} to Theorem 1.

Consider the three-qubit bit-flip map, used as the example in \cite{shabani_2009},
\begin{align}\label{eq:exLidar1}
\Phi(\rho)=c_0\rho+c_1\sum_{n=1}^3X_n\rho X_n,
\end{align}
where $X_n$ is the $\sigma_x$ Pauli matrix acting on the $n^\text{th}$ qubit, and $c_0$ and $c_1$ are real, have opposite sign, and $c_0+3c_1=1$. The corresponding CP map is $\tilde{\Phi}_{\text{CP}}(\rho)=\abs{c_0}\rho+\abs{c_1}\sum_{n=1}^3X_n\rho X_n$, the code space is $\mathcal{C}=\text{span}\{\ket{000},\ket{111}\}$, and the projector onto the code space is $P=\op{000}+\op{111}$. Then,
\begin{align}\label{eq:exLidar2}
\mathcal{R}_{\text{CP}}\left[\Phi(P\rho P)\right]\propto P\rho P,
\end{align}
where $\mathcal{R}_{\text{CP}}$ (given below) is the CP recovery map for $\tilde{\Phi}_{\text{CP}}$. However, it turns out that the code space is not in the domain of the error map \eqref{eq:exLidar1} and thus performing \eqref{eq:exLidar2} leads to negative probabilities. This is what our Corollary \ref{coll:ncpec0} predicts. 

Let 
\begin{align}\label{eq:exLidar3}
\notag\rho=&a\op{000}+(1-a)\op{111}\\
&+c^*\op{111}{000}+c\op{000}{111},
\end{align}
be a valid arbitrary density matrix in the code space. 

Applying the error map \eqref{eq:exLidar1} onto \eqref{eq:exLidar3}, we get
\begin{align}\label{eq:exLidar4}
\notag \Phi(\rho)=&c_0[a\op{000}+(1-a)\op{111}\\
\notag &+c^*\op{111}{000}+c\op{000}{111}]\\
\notag &+c_1[a\op{100}+(1-a)\op{011}\\
\notag &+c^*\op{011}{100}+c\op{100}{011}]\\
\notag &+c_1[a\op{010}+(1-a)\op{101}\\
\notag &+c^*\op{101}{010}+c\op{010}{101}]\\
\notag &+c_1[a\op{001}+(1-a)\op{110}\\
\notag &+c^*\op{110}{001}+c\op{001}{110}]\\
&=\rho'
\end{align}
If we measure with projectors in the computational basis, the probabilities are given by $tr(P_j\rho')$. Then, the $\op{000}$, $\op{111}$, $\op{100}$, and $\op{011}$ outcomes have corresponding probabilities $tr(\op{000}\rho')=c_0a$, $tr(\op{111}\rho')=c_0(1-a)$,
$tr(\op{100}\rho')=c_1a$, and $tr(\op{011}\rho')=c_1(1-a)$. If $\rho'$ is a valid density matrix then it should be positive semidefinite, and these values should be greater than or equal to zero. Here, regardless of which $c_i$ is negative [as in the definition of $\Phi(\rho)$], one of the resulting probabilities is negative. Thus, $\rho'$ is not positive semidefinite and the code space $\mathcal{C}=\text{span}\{\ket{000},\ket{111}\}$ is not in the domain of $\Phi(\rho)$.  

\begin{remark} We should emphasize here that we assume that the recovery map occurs after the error map so the output of the error map needs to be valid. However, it may be possible to implement the error and recovery maps together. In the latter situation, the code space would not need to be in the domain of the error map. 

The recovery map for $\tilde{\Phi}_{\text{CP}}(\rho)$ is
\begin{align}
\mathcal{R}_\text{CP}(\rho)=P\rho P+\sum_{n=1}^3 PX_n\rho X_nP
\end{align}
If we apply this recovery to \eqref{eq:exLidar4}, for states in the code space we see that we get back to the initial state $P\rho P$.  One may suppose that this works on average, but the processes of measurement, followed by a recovery, are nonphysical.  
\end{remark}

\begin{remark} 
Shabani and Lidar \cite{shabani_2009} consider Hermitian maps to be physical.  The negativity of the outcome is not regarded.  For example, later in \textit{Corollary 2}, they consider a Hermitian recovery without regard to its positivity \cite{shabani_2009}.
\end{remark}


\section{summary/discussion}

In this paper, we address the reversibility of quantum operations for the evolution of a subsystem that does not correspond to a completely positive map.  Some researchers \cite{Pechukas_1994, Shaji_Sudarshan_2005, Rodriguez-Rosario_2008, modi2012PositivityInThePresOfInitSECorr} suppose this is possible for a system that is initially correlated with its environment. The effects and reversibility of these more general error models (i.e., NCP errors) were considered by both Aharonov and Ben-Or \cite{aharonov2008FaultTolQuantComp}, and also Shabani and Lidar \cite{shabani_2009}. In this paper, the map describing the evolution can be physical or nonphysical. The results apply to both cases.

In general, we find that there are restrictions on the applicability of the standard quantum error correcting code conditions for evolutions that are not describable by a CP map if one is to expect a positive outcome for the operators.  These restrictions are described in our Theorem \ref{thm:nogo0} that shows that the diagonal CP error correcting conditions can fail to give a code space that has a positive output for these evolutions.  

In Corollary \ref{coll:ncpec0}, we also showed that when the pseudounitary degree of freedom diagonalizes the NCP error correcting conditions and the operators in the diagonalized error map are unitaries, then the code space is not in the domain of the error map in the sense that it is not positive. This implies that the quantum error correcting conditions for linear maps given in \cite{shabani_2009} must be supplemented to guarantee a positive density matrix.  

We then presented a set of sufficient conditions in Theorem \ref{thm:suff} for the reversibility of NCP errors.  This was followed by examples in \ref{sec:examples}.  In the near future, we will present other conditions for reversing NCP errors.  Correcting these types of errors may be important in systems where an uncorrelated initial system and environment state cannot be prepared.

Finally, we note that approximate QEC codes were introduced by Leung et al. \cite{leung1997ApproxQEC}. In some cases, this can lead to better codes for a particular set of errors.  In general, approximate quantum error correction schemes do not recover the initial state exactly, but high-fidelity recovered states are achievable. 
B\'eny and Oreshkov \cite{beny2010ApproxQEC} provided necessary and sufficient conditions for these approximate error correction codes to achieve a high fidelity.  Still, there are multiple measures expressing the performance of approximate codes, and fidelity might not be the best measure \cite{sainz2008QECesd, cafaro2014AppoxQEC}. In this paper, we have focused on the exact recovery of a quantum state.  However, in our Theorem \ref{thm:suff} it may be possible to approximately recover the state, that is, with high fidelity, when the negative part of the evolution is small, but nonzero.  We leave a more thorough discussion of approximate QEC protocols for future work.  

\section{Acknowledgments}
Funding for this research was provided by the National Science Foundation (NSF), MPS, PHY Award No. 1820870.  The authors thank Daniel Lidar, Alireza Shabani, and Purva Thakre for many helpful discussions.  Mark Byrd thanks the Center for Engineered Quantum Systems at Macquarie University in Sidney, Australia for funding and the members Gavin Brennen, Alexei Gilchrist, Daniel Terno, and especially Jason Twamley for helpful discussions.

\appendix

\section{Properties of Pseudo-Hermitian Matrices}\label{sec:appendix1}

\begin{definition}  A matrix is {\it pseudounitary} (PU) if
\begin{equation}\label{eq:PUform}
U^\dagger = \eta U^{-1} \eta^{-1}, 
\end{equation} 
where $\eta$ is a Hermitian matrix.  
\end{definition}

\begin{definition}  A matrix is {\it pseudo-Hermitian} (PH) if
\begin{equation}\label{eq:PHform}
H^\dagger = \eta H \eta^{-1}. 
\end{equation} 
\end{definition}

\begin{lemma}A PU matrix can be obtained from the exponentiation of a PH
  matrix.  
\end{lemma}
\begin{proof}
Let $U = \exp(-iHt)$ with $H$ PH.  Then 
\begin{align}
\notag U^\dagger= \exp(iH^\dagger t) =\exp(i\eta H \eta^{-1} t) &= \eta \exp(iHt) \eta^{-1}\\ 
&= \eta U^{-1} \eta^{-1}.
\end{align}
\end{proof}

\begin{lemma} 
\label{lemmaPUfromPH}
If a PU matrix is obtained from a matrix $H$ via $\exp(-iHt)$, then $H$ is PH.
\end{lemma}
\begin{proof}
To see this, consider that 
$$
U^{-1}U =\mathbb{I} = \eta^{-1} U^\dagger \eta U, 
$$
so letting $U=\exp(-itH)$,
$$
\left. \frac{d}{dt}U^{-1}U\right\vert_{t=0}=\eta^{-1} iH^\dagger\eta -iH,
$$
so $H$ is PH.  
\end{proof}

As mentioned in the text, in this article, the $\eta$ considered is of the form 
$\eta = \mbox{diag}(1,1,1,...,1,-1,-1,...,-1)$, with $p$ ones and $q$ minus ones.  In this case, the form of the PH is 
\begin{equation}
\left(\begin{array}{cc} A & B \\ 
-B^\dagger & C \end{array}\right),
\end{equation}
where $A$ is a $p\times p$ Hermitian matrix, $C$ is a $q\times q$ Hermitian matrix and $B$ is an arbitrary $p \times q$ matrix.  Note that this implies $\eta H$ is Hermitian if $H$ is PH and $\eta M$ is PH, if $M$ is Hermitian.  

From the definition of a pseudounitary matrix, $U\eta U^\dagger = \eta,$ and for this unitary, the signature of the matrix $\eta$ corresponds to the form of the unitary which is often denoted $U(p,q)$ to emphasize this relationship to $\eta$ with the given``signature" $p,q$.  

\begin{lemma}
\label{lemma:PHdiagbyPU}
A pseudo-Hermitian matrix $H$ is diagonalizable by a matrix $S$ via $S^{-1}HS=H_d$, where $H_d$ is diagonal and $S$ can be chosen pseudounitary.  
\end{lemma}  
The following proof is adapted from Ref.~\cite{Haber_notes}  for the diagonalization of Hermitian matrices.  
\begin{proof}
Let $v_1$ be an eigenvector of $H$.  (Every matrix has at least one eigenvector.)  Let $\lambda_1$ be its corresponding eigenvalue.  Then 
$$
Hv_1 =\lambda_1 v_1.
$$
Now we want to build a PU matrix that will diagonalize $H$.  Let $v_1$ be the first column of such a matrix and write
\begin{equation}
U = \left(\begin{array}{c|c}
			v_1 &  Y 
				\end{array}\right),
\end{equation}
where $Y$ is an $n\times (n-1)$ matrix and $v$ is an $n\times 1$ column vector.  The matrix $Y$ can be written as a set of $n\times 1$ column vectors $v_i$, $i=2,...,n$.  These vectors can be chosen orthogonal (under the $\eta$ inner product) to $v_1$.  (Or one could imagine using a Gram-Schmidt type process to make them orthogonal to $v_1$.)  Thus, 
\begin{equation}
\langle\langle v_j, v_1\rangle\rangle_\eta \equiv (v_j,\eta v_1) \equiv \sum_{k=1}^n (v_j^*)_k \eta_k (v_1)_k = 0,
\end{equation}
where $\eta_k$ are the diagonal elements of $\eta$.  (This could also be written using $\eta_{ik} = \eta_k\delta_{ik}$ and recall that $\eta_i=\pm 1$.)
This implies that 
\begin{equation}
(Y^\dagger \eta v_1)_j = \sum_{k=1}^n Y_{kj}^* \eta_k (v_1)_k = 0.
\end{equation}
This is true for each $j$, so $Y^\dagger \eta v_1 = 0$ as is $\eta Y^\dagger \eta v_1=0$.  

Now compute the following product:
\begin{align}
U^\dagger \eta H U 
&= \left(\begin{array}{c}
			v_1^\dagger \\ 
			\hline
			Y^\dagger 
		 \end{array}
    \right) 
			\eta H
	\left(\begin{array}{c|c}
			v_1 &  Y 	\end{array}\right) \nonumber \\
&= \left(\begin{array}{c|c}
		v_1^\dagger \eta H v_1 & v_1^\dagger \eta H Y \\ \hline
		 Y^\dagger \eta H v_1   & Y^\dagger\eta H Y \end{array}\right).
\end{align}
This matrix has the following structure:
\begin{equation}
\left(\begin{array}{c|c}
		1\times 1 & 1\times (n-1) \\ \hline
		 (n-1)\times 1  & (n-1)\times(n-1) \end{array}\right).
\end{equation}
Now note that the upper-left block and lower-left block are 
\begin{align}
v_1^\dagger \eta H v_1 &= \lambda_1 v_1^\dagger \eta v_1 = \pm \lambda_1 \nonumber \\
Y^\dagger \eta H v_1 &= \lambda_1 Y^\dagger \eta v_1 = 0. 
\end{align}
Now we have 
\begin{equation}
U^\dagger \eta H U = \left(\begin{array}{c|c}
		\pm \lambda_1 & v_1^\dagger \eta H Y \\ \hline
		     0   & Y^\dagger\eta H Y \end{array}\right).
\end{equation}
Recall that $H$ is PH, so $H=\eta H^\dagger \eta$.  This implies that 
\begin{align}
(v_1^\dagger \eta H Y)^\dagger &= Y^\dagger H^\dagger \eta v_1 
  = Y^\dagger \eta \eta H^\dagger \eta v_1 \nonumber \\
    &= Y^\dagger \eta H v_1 
    = \lambda_1 Y^\dagger \eta v_1 =0,
\end{align}
where we have used the fact that $\eta^2=1$ and $\eta^\dagger = \eta$.  
At this point, we have that 
\begin{equation}
U^\dagger \eta H U = \left(\begin{array}{c|c}
		\pm \lambda_1 & 0 \\ \hline
		     0   & Y^\dagger\eta H Y \end{array}\right).
\end{equation}

Now note that $N\equiv \eta U^\dagger \eta H U$ is PH since 
\begin{align}
\eta N^\dagger \eta &= \eta (\eta U^\dagger \eta H U)^\dagger \eta 
          = \eta (U^\dagger H^\dagger \eta U \eta ) \eta \nonumber \\
          &= \eta U^\dagger \eta (\eta H^\dagger \eta ) U 
          = \eta U^\dagger \eta H U = N,
\end{align}
where we have again used $\eta^2=1$ and $\eta^\dagger = \eta$.  Also, note that $N$ clearly has the same form as $U^\dagger \eta H U$.  

To see that $N$ has the same eigenvalues as $H$, we need only notice that $\eta U^\dagger \eta = U^{-1}$ implies this since
\begin{equation}
U^{-1} H U w = \lambda w \Rightarrow H U w = \lambda U w. 
\end{equation}
So letting $Uw = v$, we see that for any $\lambda$ 
\begin{equation}
H v = \lambda v
\end{equation}
Thus, the eigenvalues of $N$ are the same as those of $H$.  Notice that $\eta U^\dagger \eta = U^{-1}$ is exactly the PU condition that $U^\dagger \eta U = \eta$.  

Given the form of the PH matrix, Eq.~(\ref{eq:PHform}), the matrix $\eta^\prime Y^\dagger \eta H Y$ is also PH, where $\eta^\prime$ is the same as 
$\eta$, albeit with one less diagonal entry.  Thus, since $p$ and $q$ were arbitrary, this matrix can be treated in the exact same way as $H$.  We can find an eigenvector and eigenvalue and reduce it in size by $1$, leaving another PH matrix as a submatrix to be diagonalized.  Continuing this allows the matrix to be diagonalized and the diagonalizing matrix is PU since $U^{-1} H U = \eta U^\dagger \eta H U$ is diagonalized.  

\end{proof}


\section{Pseudounitary Freedom in the Operator-Sum Representation}\label{sec:appendix2}

The unitary degree of freedom for operators and for the operator-sum 
representation (OSR) is useful for a variety of reasons.  The extension of the unitary freedom for positive operators is extended to operators with negative eigenvalues.  It is then shown that the freedom is also present in the OSR.



\subsection{Unitary and Pseudounitary Freedom for Operators}

The unitary degree of freedom for operators is quite important since it shows 
that there are many different decompositions of a mixed-state density operator \cite{Schrodinger:HJW}.   
This is discussed, for example, in textbooks \cite{Nielsen_Chuang_Textbook_2011,Peres:book}.  References  \cite{Mermin:HJW,Kirkpatrick:HJW} also provide interesting discussions and references.  The nonuniqueness of a mixed-state decomposition means that there are 
many different physical systems that could give rise to the same density 
operator (matrix).



The following is adapted from Nielsen and Chuang \cite{Nielsen_Chuang_Textbook_2011} with their theorem stated below. Consider a density operator 
\beq
\rho = \sum_i p_i \ket{\psi_i}\bra{\psi_i} = \sum_i \kett{\tilde{\psi}_i}\brat{\tilde{\psi}_i},
\eeq
where we define the unnormalized quantum state $\kett{\tilde{\psi}_i} \equiv \sqrt{p_i}\ket{\psi_i}$ and another decomposition of the same quantum state
\beq
\rho = \sum_j q_j \ket{\phi_j}\bra{\phi_j} = \sum_j \kett{\tilde{\phi}_j}\brat{\tilde{\phi}_j},
\eeq
where $\kett{\tilde{\phi}_j} \equiv \sqrt{q_j}\kett{\phi_j}$.  

\begin{theorem} (As stated in \cite{Nielsen_Chuang_Textbook_2011}.  It is also proven there.)
The sets $\left\{\kett{\tilde{\psi}_i}\right\}$ and $\left\{\kett{\tilde{\phi}_j}\right\}$ generate the same density matrix if and only if 
\beq
\kett{\tilde{\psi}_i} = \sum_j u_{ij}\kett{\tilde{\phi}_j},
\eeq
where $(u_{ij})$ is a unitary matrix, and we add zero vectors to the smaller set so that the two sets have the same number of elements.
\end{theorem}

Now let us suppose that our operator can be expanded in a basis $\{\ket{v_i}\}$ and a 
set of eigenvalues that are not necessarily positive, but are real, $\mu_i\in \mathbb{R}$,
\beq
\tau = \sum_i \mu_i \ket{v_i}\bra{v_i}.
\eeq
Furthermore, suppose that there is another decomposition of $\tau$ in terms of a 
set of eigenvectors $\{w_j\}$ and eigenvalues $\nu_j$ so that we also have 
\beq
\tau = \sum_j \nu_j\op{w_j}{w_j}.
\eeq
As before, we define $\ket{\tilde{v}_i}\equiv\sqrt{\mu_i}\ket{v_i}$ and $\ket{\tilde{w}_j}\equiv\sqrt{\nu_j}\ket{w_j}$.  We will also define $\eta_i \equiv$sgn$(\mu_i)$ and $\zeta_j\equiv$sgn$(\nu_j)$ to be the sign (magnitude $1$) of the eigenvalues.  Thus, $\eta_i = \pm1$; it is $+1$ for a positive eigenvalue and $-1$ for a negative eigenvalue.  

\begin{theorem}
The sets $\left\{\ket{\tilde{v}_i}\right\}$ and $\left\{\ket{\tilde{w}_j}\right\}$ generate the same operator if and only if 
\beq
\ket{\tilde{v}_i} = \sum_j u_{ij}\ket{\tilde{w}_j},
\eeq
where $(u_{ij})$ is a pseudounitary matrix, and we add zero vectors to the smaller set so that the two sets have the same number of elements.
\end{theorem}
\begin{proof}
($\Leftarrow$) 
Suppose $\ket{\tilde{v}_i} = \sum_j u_{ij}\ket{\tilde{w}_j}$, where $(u_{ij})$ is a PU matrix.  A number of zero vectors will be added to the smaller set to make them the same size.  Therefore, the two sets $\eta_i$ and $\zeta_j$ can also be made the same size.  Let us call them both $\eta_i$.  Then 
\beq
\tau = \sum_i \eta_i \ket{\tilde{v}_i}\bra{\tilde{v}_i} = \sum_{ijk} \eta_i u_{ij}\ket{\tilde{w}_j} \bra{\tilde{w}_k}u_{ik}^*,  
\eeq
and since $u_{ij}$ is PU, $\sum_i u_{ij}\eta_i u_{ik}^* = \delta_{jk}\eta_k$.  Therefore, 
\begin{align}
\notag\tau = \sum_i \eta_i \ket{\tilde{v}_i}\bra{\tilde{v}_i} &= \sum_{ijk} \eta_k \delta_{jk} \ket{\tilde{w}_j} \bra{\tilde{w}_k}\\ 
&= \sum_k \eta_k \ket{\tilde{w}_k}\bra{\tilde{w}_k}.  
\end{align}

($\Rightarrow$)
Now suppose 
\beq
\tau = \sum_i \mu_i \ket{v_i}\bra{v_i} = \sum_j \nu_j\op{w_j}{w_j}.
\eeq
Let $\tau = \sum_r \beta_r \op{\tilde{r}}{\tilde{r}}$ be another decomposition 
of $\tau$ with $\{\ket{\tilde{r}}\}$ a complete set of un-normalized orthogonal states and 
$\beta_r=\pm 1$.  The set of $\{\ket{\tilde{r}}\}$ is complete, so we can append 
zeroes to the set $\{\ket{v_i}\}$ and can take $(\beta_r)=(\eta_i)$.  Also, since the set 
$\{\ket{\tilde{r}}\}$ is complete, we can expand any $\ket{\tilde{v}_i}$ as 
\beq
\ket{\tilde{v}_i} = M_{ir}\ket{\tilde{r}}.  
\eeq
Now, since these are both decompositions of $\tau$, we have 

\beq
\tau = \sum_i \eta_i \ket{\tilde{v}_i}\bra{\tilde{v}_i}
      = \sum_{irs}\eta_i M_{ir} M_{is }^*\ket{\tilde{r}}\bra{\tilde{s}} 
      = \sum_r \eta_r \ket{\tilde{r}}\bra{\tilde{r}},
\eeq
which is true if $\eta_i M_{ir} M_{is }^* = \eta_r\delta_{rs}$.  This is just the condition for $M$ to be pseudounitary.  

Now, we could make the same argument for the decomposition in terms of $\ket{\tilde{w}_j}$.  Then, since these are each related by a PU and the 
composition of two PU matrices is a PU matrix, there exists a PU matrix that 
takes $\ket{\tilde{v}_i}$ to $\ket{\tilde{w}_j}$.    
\end{proof}


\subsection{Unitary and Pseudounitary Freedom in the OSR}

The description of the dynamical map is not unique.  It can be
represented by the set of $C_k$ corresponding to the
eigenvector decomposition of the map $B$, but there are many other 
representations.  In this section, we find an equivalence class of maps
and provide an expression of such a 
freedom after reviewing the case for completely positive maps. 

For completely positive maps, we reiterate that a theorem describing
the freedom, examples, and uses can be found in 
Ref.~\cite{Nielsen_Chuang_Textbook_2011}.


\subsubsection{Unitary Freedom for Completely Positive Maps}

Let us first quote Nielsen and Chuang \cite{Nielsen_Chuang_Textbook_2011}: 
\begin{quote}
Suppose $\{E_1, ..., E_m\}$ and
$\{F_1, ...,F_n\}$ are operation elements giving rise to quantum
operations ${\cal E}$ and ${\cal F}$ respectively.  By appending zero
operators to the shorter list of operation elements we may ensure that
$m=n$.  Then ${\cal E} = {\cal F}$ if and only if there exist complex
numbers $u_{ij}$ such that $E_i =\sum_ju_{ij}F_j$, and $u_{ij}$ is an
$m$ by $m$ unitary matrix.  
\end{quote}

Note that zero may be added to the map ${\cal F}$ in such a way that it is {\it not} obtainable from the map ${\cal E}$ by a unitary transformation.  Let us consider the following example.  Let ${\cal E} \rightarrow {\cal E}^\prime = \sum_iE_iE_i^\dagger + A A^\dagger - A A^\dagger$.  Suppose $A$ is linearly independent of all $E_i$, then this map can ${\it not}$ be obtained from the set with a unitary transformation.  The map ${\cal E}^\prime$ differs from ${\cal E}$ in some sense trivially and in practice it is very often easy to spot such ``an extension by zero.''  However, the difference could be difficult to recognize and provides a technical point to note about the theorem.  

When considering such cases, one may define an equivalence class 
of maps by identifying all maps which differ by such trivial 
extensions.  Thus, maps which are in the same class are those which
differ by the addition of operation elements which would cancel.  
The representative will always be the element of the class that has no such trivial extension.  This will be termed a base map.  
\begin{definition}
For a given equivalence class of maps which differ by a trivial
extension, the {\bf base map} of the class is the representative
of that class which has not been trivially extended.  
\end{definition}
Different base maps belong to different classes.  


\subsubsection{Pseudounitary freedom for Hermiticity-preserving maps}

Now let us consider a map $\Phi(\rho) = \sum_j \eta_j  C_j
\rho C_j^\dagger$ and introduce a set of operators $D_j$ corresponding
to another base 
map $\Phi '(\rho) = \sum_j \eta_j D_j \rho D_j^\dagger$.  As
stated above, we may take $\eta_j = \pm 1$.  We can choose 
the number of operators to be the same by appending
zero operators to the shorter list.  
This enables the number of $-1$ and $+1$ to be chosen to be the same
for each of the maps.  Furthermore, we will order the
set of $\eta_j$ such that the first $p$ are $+1$ and the next $q$ are
$-1$.  

The freedom in the operator-sum representation 
is described by the group $U(p,q)$.  This
group is often called a pseudounitary group due to its relation to the
unitary group and it is a metric-preserving group with the signature
of the metric determined by the integers $p,q$.  See
for example (\cite{Gilmore:book}, pages 45, 197),
(\cite{Cornwell:84II}, page 392), (\cite{Wybourne}, page 12), or 
(\cite{Helgason:DS}, page 444).  

Let $\eta$ be an $N\times N$ diagonal matrix with
the first $p$ entries $+1$, the next $q$ entries $-1$, and $N=p+q$.  Then
for all $U\in U(p,q)$, 
\begin{equation}
\label{eq:upqcond}
U^\dagger \eta = \eta U^{-1}.
\end{equation}
We may express the matrix $\eta$ as a diagonal matrix with the matrix
elements being $\eta_k$, $\eta_k = +1$, for $k = 1, ..., p$ and
$\eta_k = -1$, for $k=p+1, ..., p+q=N$.  Alternatively, we may express
the matrix $\eta$ using elements $(\eta)_{kl} = \eta_k \delta_{kl}$.
This is a diagonal matrix with the first $p$
entries along the diagonal are $+1$ and the next $q$ are $-1$.  Let the
elements of the matrix $U$ be given by $u_{ij}$ and those of 
$U^\dagger$ be $u_{ji}^*$.  Then Eq.~(\ref{eq:upqcond}) can be
written as $U^\dagger \eta U = \eta$, or since $\eta^2 = \mathbb{I}$, 
$U \eta U^\dagger = \eta$.  In components, this can be written as 
\begin{equation}
\sum_{jk} u_{ij} \eta_j \delta_{jk} u_{lk}^* = \eta_i\delta_{il}. 
\end{equation}
Having established this property for elements of the group $U(p,q)$,
the following theorem may now be stated and proved.  (Originally, a version of the following proof was presented in Ref.~\cite{Ou_Byrd_2010}.)

\begin{theorem}{Pseudounitary freedom:}
Suppose $\{C_1,C_2, ..., C_n\}$ and  $\{D_1,D_2, ..., D_m\}$, 
are operation elements giving rise to base 
quantum operations (maps) 
$\Phi$ and $\Phi '$ respectively.  Explicitly, 
\begin{equation}\label{eq:CandDeqs}
\Phi = \sum_i\gamma_iC_iC_i^\dagger,
\;\;\; \mbox{and} \;\;\;
\Phi^\prime = \sum_j \mu_j D_jD_j^\dagger,
\end{equation}
where each $\gamma_i$ and each $\mu_j$ is $\pm 1$ and ordered as
above, with all $+1$ eigenvalues first.  Furthermore, we can always
take $\gamma_i = \mu_i$ with zero-valued $C_i$ or zero-valued $D_j$
appended to the shorter list for the $+1$ $(-1)$ eigenvalue.  
Then $\Phi=\Phi^\prime$ if and only if 
\begin{equation} \label{e123}
D_j =\sum_{i} u_{ji}C_{i},
\end{equation}
where the numbers $u_{ji}$ form a $p+q$ by $p+q$ 
matrix in $U(p,q)$.  
\end{theorem}

{\it Proof:} We first consider whether the condition is necessary and
use the notation $C_i=\ket{i}$, $D_i=\ket{j}$.  Suppose that 
\begin{equation}
\Phi =\Phi^\prime.
\end{equation}
[Or, if one would like to display the argument explicitly, 
$\Phi(\rho) =\Phi^\prime(\rho)$.]  For a general map $\Phi$, there
exists a corresponding $B$ matrix such
that $\Phi=B$ [i.e., $\Phi (\rho) = B \rho$].  $B$ has an eigenvector
decomposition $B=\sum_{k^\prime} \lambda_k^\prime
\ket{k^\prime}\bra{k^\prime}$ where the set of
$\ket{k^\prime}$ are linearly independent since they are orthogonal. 
This follows
from the fact that the eigenvectors can be chosen orthogonal. Now 
$|k\rangle=\sqrt{|\lambda_k^\prime|}\;
|k^\prime\rangle$.  These vectors are clearly also orthogonal and thus
linearly independent if the $\ket{k^\prime}$ are.  Then $B$ can be
re-expressed as 
$B=\sum_k \eta_k \ket{k}\bra{k}$ with the first $p$ eigenvalues 
$\eta_k=+1, \; k= 1,...,p$ and the next 
$q$ eigenvalues $\eta_k=-1\;, k=p+1,...,p+q$.  This gives 
\begin {equation}
B =\sum_k \eta_k \ket{k}\bra{k} 
  = \sum_{k=1}^{p} \ket{k}\bra{k}-\sum_{k=p+1}^{p+q} \ket{k}\bra{k},
\end{equation}
which is an eigenvector decomposition of the map $\Phi$.  
Now, let us consider another decomposition of $B$ corresponding to the
set of $C_i$, $B=\sum_{i} \gamma_i \ket{i}\bra{i}$.  Each $|i\rangle$
can be written as a linear combination of the $|k\rangle$,
$|{i}\rangle =\sum_{k} w_{ik}|{k}\rangle$. (See, for example, 
Ref.~\cite{Nielsen_Chuang_Textbook_2011}, page 104.) 
Given $\Phi=B$, 
\begin{equation}
  \sum_k \eta_k  \ket{k}\bra{k} = \sum_{kl}\left(\sum_i 
                \gamma_i w_{ik}w^*_{il}\right)\ket{k}\bra{l}.  
\end{equation}
Since the $\ket{k}$ are linearly independent, it is clear that this
can only happen if 
\begin{equation} \label {e1}
\sum_i\gamma_i w_{ik}w^*_{il} = \delta_{kl}\eta_k.
\end{equation}
We may always take $\eta_i = \gamma_i$ by appending the shorter list
of vectors ($\{\ket{i}\}$ or $\{\ket{k}\}$) with zero vectors.  
This will ensure the matrices $\gamma$ with elements
$\delta_{ij}\gamma_i$ and $\eta$ with elements $\delta_{kj}\eta_k$ are
equal.  Furthermore,  $w$ can then be taken to be square with $\ket{i}
= \sum_k w_{ik}\ket{k}$.  The condition, Eq.~(\ref{e1}), can then be
written as 
\begin{equation}
w^\dagger \eta w = \eta,
\end{equation}
which is the condition for the matrix $w$ to be in $U(p,q)$.  
Now we can use the same argument with $B=\Phi^\prime$ and 
$v_{jk}$ such that  $\ket{j} = \sum_k v_{jk}\ket{k}$ to show
\begin{equation} \label{a7}
v^\dagger \eta v =\eta. 
\end{equation}
Since each of these two are related to the same expression for $B$
using elements of $U(p,q)$ which is a group, then the linear
transformation which takes the $C_i$ to the $D_j$ is given by 
$u=vw^{-1}$ and is in $U(p,q)$.  

Next we consider whether $u\in U(p,q)$ will imply that
$\Phi=\Phi^\prime$, i.e., if the condition is sufficient.  
This is straightforward algebra. 
Then Eq. \eqref{eq:CandDeqs} is
\begin{eqnarray} \nonumber
  \Phi^\prime(\rho) &=& \sum_j \mu_j D_j \rho D_j^{\dag} 
  = \sum_{lkj} \mu_j u_{jl}u^*_{jk}C_l \rho C_k^{\dag}\\ \nonumber
   &=& \sum_{lk}\left(\sum_j \mu_ju_{jl}u^*_{jk}\right)C_l \rho
   C_{k}^{\dag} \\ \nonumber
     &=& \sum_{lk}\gamma_l \delta_{lk} C_l \rho C_{k}^{\dag} 
            = \Phi (\rho),
\end{eqnarray}
which shows that the two sets of operators $C_j$ and $D_j$ related 
by a pseudounitary matrix $u$ will yield the same map. $\square$


%

\end{document}